\documentclass{revtex4-1}%

\usepackage{epsfig}

\usepackage{amssymb}

\usepackage{xcolor}

\usepackage[english]{babel}

\begin{document}

\title{Photonic Crystal-Based Compact High-Power Vacuum Electronic Devices}

\author{V.G. Baryshevsky, A.A. Gurinovich}

\address{Research Institute for Nuclear Problems, Minsk, Belarus}

\begin{abstract}
This paper considers how the finite dimensions of a photonic
crystal placed inside a resonator or waveguide affect the law of
electron beam instability.
The dispersion equations describing e-beam instability in the
finite photonic crystal placed inside the resonator or waveguide
(a bounded photonic crystal) are obtained.
Two cases are considered: the conventionally considered case, when
diffraction is suppressed, and the case of direct and diffracted
waves having almost equal amplitudes. The instability law is shown
to be responsible for increase of increment of instability and
decrease of length, at which instability develops, for the case
when amplitude of diffracted wave is comparable with that of
direct one, that happens in the vicinity of $\pi$-point of
dispersion curve. Application of photonic crystals for development
of THz sources at electron beam current densities available at
modern accelerators is discussed.

\end{abstract}

\maketitle

\section{Introduction}

THz sources are in demand for a variety of applications:
information and communications technology, biology and medicine,
non-destructive investigations and homeland security, food and
agricultural products quality control, global environmental
monitoring, space research and ultra-fast computing.
High-power tunable THz sources are very important devices to bring
promising prospects to a wide use; high efficiency and compactness
are also highly desired (see \cite{THZ-review1,THZ-review2} and
references therein).

Sources of THz radiation are studied by numerous authors through
expansion to THz range of approaches  and principles, which
generally serve for microwave sources, namely:  travelling wave
tubes, backward wave oscillators,  generators utilizing
diffraction or Smith-Purcell radiation, etc
\cite{poster_1,poster_2,12,poster_4}.
A slow-wave structure (SWS) is generally used in such devices to
make the electromagnetic wave phase velocity less than the speed
of light so that beam electrons can interact with the wave and
convert their energy to radiation.
A diffraction grating, a helical line, a corrugated waveguide, a
photonic crystal, a multipin structure 
 or a spatially periodic
structure of any type, they all could work as SWS and enable
frequency tuning by change of their geometry
\cite{10,NIM2005,9,9+,tuning1}.

The general feature for all the above listed radiation sources is
the instability of an electron beam in a spatially periodic media,
which results in beam self-modulation and radiation of
electromagnetic waves.
The increment $\delta_0 \sim Im k_z$
($k_z$ is the longitudinal wave number),
which describes the electron beam instability responsible for
radiation process in such devices, conventionally is determined by
the
{unperturbed density $\rho_{b0}$ of electrons in the beam as
follows: $\delta_0 \sim \sqrt[3]{\rho_{b0}}$}
\cite{Miller,Granatstein,Tsimring}. The threshold current density
$j_{thr}$, which is required for coherent electromagnetic
oscillations to grow, in this case depends on the beam-wave
interaction length $L$ as $j_{thr} \sim L^{-3}$.

Shift to THz range for conventional microwave devices faces
several difficulties limiting the output power due to drastic
decrease in dimensions of the interaction area. SWS period tends
to submillimeter range and requirements to electron beam quality
and guiding precision become more strict. The amplitude of the
harmonic, which is in synchronism with the beam, decreases with
the distance from the SWS surface on the scale
\begin{equation}
\Delta=\frac {\lambda \beta \gamma}{2 \pi},
\end{equation}
where $\lambda$ is the wavelength, $\beta=\frac{v}{c}$, $\gamma$
is the Lorentz factor, $v$ and $c$ are the speed of electron beam
and light, respectively.
Thus, only electrons moving close to the structure can efficiently
interact with the wave; for example, $\delta \approx 0.1$\,mm for
100\,keV electron beam and $\lambda=$1\,mm.
Small dimensions constrain applicable electron beam currents and,
therefore, available output power.
Therefore, approaches enabling increase of efficiency and
transverse dimensions of interaction area (electron-beam cross
section) are of high priority for THz source development.
Such approaches were for the first time ever proposed for X-ray
range by the concept of volume free electron laser (VFEL)
\cite{Ch_1,9,Ch_6,Ch_7}, which enables increase of both the
efficiency and the transverse coherence area, and simultaneous
reduction of threshold current density and operation length.
Successive expansion of VFEL concept to microwave
\cite{10,NIM2005,12,poster_4,Ch_4}, terahertz
\cite{THZ_VFEL1,THZ_VFEL2} and visible light \cite{opticalVFEL}
ranges includes both theoretical and experimental studies;
diversely designed SWS are used, namely: diffraction gratings,
photonic crystals, etc.

Another law of electron beam instability was discovered  in
\cite{Ch_1,9,Ch_6,Ch_7}. It was found there that for an electron
beam moving in a crystal for the case, when Bragg diffraction
could occur,
the electron beam instability increment $\delta_0$ could turn out
to be determined by $\delta_0 \sim \sqrt[4]{\rho_{b0}}$ rather
then conventional $\delta_0 \sim \sqrt[3]{\rho_{b0}}$
\cite{Miller,Granatstein,Tsimring}.
The law of electron beam instability inherent for volume free
electron laser \cite{Ch_1,9,Ch_6,Ch_7} is revealed when
distributed feedback is formed by Bragg diffraction of
beam-induced electromagnetic wave by a spatially periodic SWS.
This law gives for the threshold current density the very
different dependance on interaction length $L$ as follows:
$j_{thr} \sim L^{-3-2s}$, where $s$ is the number of additional
waves arisen due to diffraction (when one additional wave arises
then $j_{thr} \sim L^{-5}$).

For a certain slow-wave structure this law reveals only for narrow
range of parameters (electron beam energy and radiation
frequency), for which the group velocity of the excited wave is
close to zero i.e. synchronism of beam and wave is achieved in the
vicinity of $\pi$-point on the SWS dispersion curve $k(\omega)=0$
($k$ is the wavenumber, $\omega$ is the radiation frequency).
For example, when in addition to the electromagnetic wave, which
is excited by electron beam via its interaction with SWS, one more
wave, which propagates in the direction determined by the Bragg's
law, is present, one could expect that the threshold current
density $j_{thr}$ is defined by law $j_{thr} \sim L^{-5}$ rather
than $j_{thr} \sim L^{-3}$ for proper combination of electron beam
energy and radiation frequency.
In particular, the law $j_{thr} \sim L^{-5}$ can be observed in
conventional backward wave oscillator with corrugated waveguide in
case when the excited wave and the diffracted one propagate in
opposite directions along the system axis.

First lasing of VFEL, which uses the above described instability
law, was presented in \cite{10}. Frequency tuning in this
generator is analyzed in \cite{NIM2005}.
Theoretical study of the instability of electron beams moving in
natural and artificial (photonic) crystals was carried out for the
ideal case of an infinite medium (see the review \cite{9} and
\cite{Ch_1,10,11,12,Ch_4}).
It is known that the discrete structure of the modes in waveguides
and resonators is crucial for effective generation in the
microwave range \cite{3,4,5}.
This paper considers how the finite dimensions of a photonic
crystal placed inside a resonator or waveguide affect the law of
electron beam instability.

The dispersion equation describing e-beam instability in the
finite photonic crystal placed inside the resonator or waveguide
(a bounded photonic crystal) is obtained. The instability law  is
shown to be valid and caused by mixing of the electromagnetic
field modes in the finite volume due to the periodic disturbance
produced by the photonic crystal.

\section{Equations describing motion of a relativistic electron beam
in a bounded photonic crystal}

To describe generation of induced radiation (i.e. electron beam instability) in either a photonic or a natural crystal
one should start from Maxwell equations:

\begin{eqnarray}
 rot\vec {H} = \frac{{1}}{{c}}\frac{{\partial \vec
{D}}}{{\partial t}} + \frac{{4\pi} }{{c}}\vec {j},\;rot\vec {E} =
- \frac{{1}}{{c}}\frac{{\partial \vec {H}}}{{\partial t}}, \\
 div\vec {D} = 4\pi \rho ,\;\frac{{\partial \rho}
}{{\partial t}} + div\vec {j} = 0,\nonumber
\label{eq1}
\end{eqnarray}

\noindent where $\vec {E}$ and  $\vec {H}$  are the strength of
the electric and the magnetic field, respectively;
 $\vec {j}$ and $\rho$ are the current and
 {charge} densities;
 $D_{i} \left( {\vec
{r},t^{\prime} } \right) = \int \varepsilon _{il} \left( {\vec
{r},t - t^{\prime} } \right)E_{l} \left( {\vec {r},t^{\prime} }
\right)dt^{\prime} $ or $D_{i} \left( {\vec {r},\omega}  \right) =
\varepsilon _{il} \left( {\vec {r},\omega} \right)E_{l} \left(
{\vec {r},\omega}  \right)$, where indices $i,l = 1,2,3$
correspond to $x,y,z$;  $\varepsilon _{il} \left( {\vec
{r},\omega}  \right)$ is the dielectric permittivity tensor of the
photonic crystal.

The current and charge densities are defined as:

\begin{equation}
\label{eq2} \vec {j}\left( {\vec {r},t} \right) = e \sum_\alpha
{\vec {v}}_\alpha \delta(\vec {r} -\vec {r}_\alpha(t)),
{~\rho(\vec {r},t)=e \sum_\alpha \delta(\vec {r} -\vec
{r}_\alpha(t)) = e \rho_b (\vec {r},t)\,,}
\end{equation}

\noindent where $e$ is the electron charge, $\rho _{b} $ is the
beam density (the number of electrons per 1 cm$^{3}$). The
velocity $\mathop {\vec {v}}_{\alpha}$ of electron with number
$\alpha $ reads as

\begin{equation}
\label{eq3} \frac{{d  {\vec {v}}_{\alpha} } }{{dt}} =
\frac{{e}}{{m\gamma _{\alpha} } }\left\{ {\vec {E}\left( { {\vec
{r}}_{\alpha} \left( {t} \right),t} \right) +
\frac{{1}}{{c}}\left[ {  {\vec {v}}_{\alpha} \left( {t} \right)
\times \vec {H}\left( {  {\vec {r}}_{\alpha} \left( {t} \right),t}
\right)} \right] - \frac{{  {\vec {v}}_{\alpha} } }{{c^{2}}}\left(
{  {\vec {v}}_{\alpha}  \left( {t} \right)\vec {E}\left( { {\vec
{r}}_{\alpha}  \left( {t} \right),t} \right)} \right)} \right\},
\end{equation}

\noindent where $\gamma _{\alpha}  = \left( {1 -
{\textstyle{{v_{\alpha} ^{2}}  \over {c^{2}}}}} \right)^{ -
{\textstyle{{1} \over {2}}}}$ is the Lorentz factor, $\vec
{E}\left( { {\vec {r}}_{\alpha}  \left( {t} \right),t} \right)$
($\vec {H}\left( {  {\vec {r}}_{\alpha}  \left( {t} \right),t}
\right)$) is the electric (magnetic) field strength at point $
{\vec {r}}_{\alpha}  $, where the electron with number $\alpha $
is located. Note that equation (\ref{eq3}) can be written as
follows \cite{14}:

\begin{equation}
\label{eq4} \frac{{d  {\vec {p}}_{\alpha} } }{{dt}} =
m\frac{{d\gamma _{\alpha}  \vec{v}_{\alpha} } }{{dt}} = e\left\{
{\vec {E}\left( {  {\vec {r}}_{\alpha}  \left( {t} \right),t}
\right) + \frac{{1}}{{c}}\left[ {  {\vec {v}}_{\alpha} \left( {t}
\right) \times \vec {H}\left( {  {\vec {r}}_{\alpha} \left( {t}
\right),t} \right)} \right]} \right\},
\end{equation}

\noindent
where $p_{\alpha}  $ is the particle momentum.

From equations (\ref{eq1}) one can obtain

\begin{equation}
\label{eq5}
 - \Delta \vec {E} + \vec {\nabla} \left( {\vec {\nabla} \vec {E}} \right) +
\frac{{1}}{{c^{2}}}\frac{{\partial ^{2}\vec {D}}}{{\partial t^{2}}} = -
\frac{{4\pi} }{{c^{2}}}\frac{{\partial \vec {j}}}{{\partial t}}.
\end{equation}

The dielectric permittivity tensor can be presented in the form $\hat
{\varepsilon} \left( {\vec {r}} \right) = 1 + \hat {\chi} \left( {\vec {r}}
\right)$, where $\hat {\chi} \left( {\vec {r}} \right)$ is
the susceptibility.

For $\hat {\chi}  < < 1$, equation (\ref{eq5}) can be rewritten as
follows

\begin{equation}
\label{eq6} \Delta \vec {E}\left( {\vec {r},t} \right) -
\frac{{1}}{{c^{2}}}\frac{{\partial ^{2}}}{{\partial t^{2}}}\int
\hat {\varepsilon} \left( {\vec {r},t - t^{\prime} } \right)\vec
{E}\left( {\vec {r},t^{\prime} } \right)dt^{\prime} = 4\pi \left(
{\frac{{1}}{{c^{2}}}\frac{{\partial \vec {j}\left( {\vec {r},t}
\right)}}{{\partial t}} + \vec {\nabla} \rho \left( {\vec {r},t}
\right)} \right).
\end{equation}

In the general case, the susceptibility of the photonic crystal
reads
\begin{equation}
\hat {\chi} \left( {\vec {r}} \right) = \sum\limits_{i} {\hat {\chi
}_{cell} \left( {\vec {r} - \vec {r}_{i}}  \right)} ,
\label{eq:chi}
\end{equation}
where $\hat
{\chi }_{cell} \left( {\vec {r} - \vec {r}_{i}}  \right)$ is the
susceptibility of the crystal unit cell. The susceptibility of an
infinite perfect crystal $\hat {\chi} \left( {\vec {r}} \right)$
can be expanded into a Fourier series as follows: $\hat {\chi}
\left( {\vec {r}} \right) = \sum\limits_{\vec {\tau} } {\hat
{\chi} _{\vec {\tau} } e^{i\vec {\tau} \vec {r}}} ,$ where $\vec
{\tau} $ is the reciprocal lattice vector of the crystal.

Let us consider in details a practically
important case of a bounded photonic crystal; to be more specific, let us study a photonic crystal placed inside a
 waveguide of rectangular cross-section with smooth walls.
The eigenfunctions and eigenvalues of a rectangular waveguide are
well-studied \cite{15,16}.
Suppose $z$-axis to be directed along the waveguide axis, $a$ and $b$ are the waveguide dimensions along $x$ and $y$ axes, respectively. Let's
make Fourier transform of (\ref{eq5}) over time and longitudinal
coordinate $z$. Expanding thus obtained equation for the field
$\vec {E}\left( {\vec {r}_{ \bot}  ,k_{z} ,\omega} \right)$ over a
full set of vector eigenfunctions $\vec
{Y}^{\lambda }_{mn} \left( {\vec {r}_{ \bot} ,k_{z}} \right)$ of a rectangular waveguide
(where $m,n = 1,2,3....$ and $\lambda $ describes the type of
the wave \cite{17})  one can obtain for field $\vec {E}$ the
equality

\begin{equation}
\label{eq7}
\vec {E}\left( {\vec {r}_{ \bot}  ,k_{z} ,\omega}  \right) =
\sum\limits_{mn\lambda}  {C^{\lambda} _{mn} \left( {k_{z} ,\omega}
\right)\vec {Y}^{\lambda} _{mn}}  \left( {\vec {r}_{ \bot}  ,k_{z}}
\right).
\end{equation}

As a result, the following system of equations can be written

\begin{equation}
\label{eq8}
\begin{array}{l}
 \left[ {\left( {k_{z} ^{2} + \varkappa _{mn\lambda}  ^{2}} \right) -
\frac{{\omega ^{2}}}{{c^{2}}}} \right]C^{\lambda} _{mn} \left( {k_{z}
,\omega}  \right) - \\
 - \frac{{\omega ^{2}}}{{c^{2}}}\,\frac{{1}}{{2\pi
}}\sum\limits_{m'n'\lambda '} {\int {\vec {Y}^{\lambda ^{\ast} }_{mn} \left(
{\vec {r}_{ \bot}  ,k_{z}}  \right)\hat {\chi} \left( {\vec {r}} \right)\vec
{Y}^{\lambda '}_{m'n'} \left( {\vec {r}_{ \bot}  ,k'_{z}}  \right)e^{ -
i\left( {k_{z} - k'_{z}}  \right)z}} \,} d^{2}r_{ \bot}  C_{m'n'}^{\lambda
'} \left( {k'_{z} ,\omega}  \right)dk'_{z} dz = \\
 = \frac{{4\pi i\omega} }{{c^{2}}}\int {\vec {Y}^{\lambda ^{\ast} }_{mn}
\left( {\vec {r}_{ \bot}  ,k_{z}}  \right)\left\{ {\vec {j}\left( {\vec
{r}_{ \bot}  ,z,\omega}  \right) + \frac{{c^{2}}}{{\omega ^{2}}}\vec {\nabla
}\left( {\vec {\nabla} \vec {j}\left( {\vec {r}_{ \bot}  ,z,\omega}
\right)} \right)} \right\}e^{ - ik_{z} z}} d^{2}r_{ \bot}  dz \\
 \end{array}
\end{equation}
\noindent where $\varkappa _{mn\lambda}  ^{2} = k_{xm\lambda} ^{2}
+ k_{yn\lambda} ^{2} $.

The beam current and density, those appear on the right-hand side of
(\ref{eq8}), are the complicated functions of $\vec {E}$. To
study the system instability, it is sufficient to
consider it in the approximation linear  over
perturbation, i.e., one can expand the expressions for $\vec {j}$
and $\rho $ over $\vec {E}$ amplitude and confine
oneself with the linear approximation.

As a result, a closed system of equations comes out. For further
consideration, one should find the expressions for
corrections to beam current density $\delta \vec {j}$ and beam charge density $\delta  \rho $, which arise due to beam
perturbation by the field. Considering the Fourier transforms of
current and charge densities $\vec {j}( {\vec
{k},\omega} )$ and $\rho ( {\vec {k},\omega} )$,
 one can obtain from (\ref{eq2}) that

\begin{equation}
\delta \vec {j}( {\vec {k},\omega}  ) = e\sum\limits_{\alpha =
1}^{N} {e^{ - i\vec {k}\vec {r}_{\alpha _{0}} } } \left\{ {\delta
\vec {v}_{\alpha}  \left( {\omega - \vec {k}\vec {u}_{\alpha} }
\right) + \vec {u}_{\alpha}  \frac{{\vec {k}\delta \vec
{v}_{\alpha}  \left( {\omega - \vec {k}\vec {u}_{\alpha} }
\right)}}{{\omega - \vec {k}\vec {u}_{\alpha} } }} \right\},
\end{equation}

\noindent
where $\vec {r}_{\alpha _{0}}  $ is the initial
coordinate of the electron, $\vec
{u}_{\alpha}  $ is the unperturbed velocity of the electron.

For simplicity, let us consider a cold beam, for which $\vec
{u}_{\alpha}  \approx \vec {u}$, where $\vec {u}$ is the mean
velocity of the beam. The general case of a hot beam can be obtained
by averaging $\delta \vec {j}\left({\vec {k},\omega}  \right)$
over distribution of particle the velocities $\vec {u}_{\alpha}$  in the beam.

According to (\ref{eq3}),  velocity $\delta \vec {v}_{\alpha} $
is determined by  field $\vec {E}\left( {\vec {r}_{\alpha}
,\omega}  \right)$, where $\vec
{r}_{\alpha}$ is the electron location point. The Fourier transform of $\vec {E}\left(
{\vec {r}_{\alpha}  ,\omega}  \right)$ has a form

\[
\vec {E}( {\vec {r}_{\alpha}  ,\omega}  ) = \frac{{1}}{{\left(
{2\pi}  \right)^{3}}}\int {\vec {E}( {\vec {k}',\omega} )e^{i\vec
{k}'\vec {r}_{\alpha} } d^{3}k'}.
\]

As a result,  $\delta \vec {j}( {\vec {k},\omega}
)$ includes the following sum over the
beam particles $\sum\limits_{\alpha}  {e^{ - i\left( {\vec
{k} - \vec {k}'} \right)\,\vec {r}_{\alpha} } } $. Suppose that the electrons in
an unperturbed beam are uniformly distributed over the area
occupied by the beam. Therefore

\[
\sum\limits_{\alpha}  {e^{ - i\left( {\vec {k} - \vec {k}'}
\right)\,\vec {r}_{\alpha} } } = \left( {2\pi} \right)^{3}{\rho
_{b0}} \,\delta ( {\vec {k} - \vec {k}'} ),
\]

\noindent where $\rho _{b0} $ is the unperturbed beam density (the
number of electrons per 1 cm$^{3}$).

As a result, the following expression for $\delta \vec {j}( {\vec
{k},\omega} )$ can be obtained \cite{18,19}:

\begin{equation}
\label{eq9} \delta \vec {j}({\vec {k},\omega}) =  i \frac{{\vec
{u} \,{e^{2}\rho_{b0} } \left( {k^{2} - \frac{{\omega
^{2}}}{{c^{2}}}} \right)}}{{\left( {\omega - \vec {k}\vec {u}}
\right)^{2}m\,\gamma\, \omega}\, }\, \vec {u} \, \vec {E}( {\vec
{k},\omega} ).
\end{equation}
{The expression for $\rho_b ( {\vec {k},\omega})$ one can obtain
using the continuity equation.}
Expression (\ref{eq9}), the inverse Fourier transform of $\vec
{E}( {\vec {k},\omega} )$, and the expansion
(\ref{eq7}) enable writing the system of equations (\ref{eq8}) as
follows:
\begin{equation}
\label{eq10}
\begin{array}{l}
 \left[ {\left( {k_{z} ^{2} + \varkappa _{mn\lambda}  ^{2}} \right) -
\frac{{\omega ^{2}}}{{c^{2}}}} \right]C^{\lambda} _{mn} \left( {k_{z}
,\omega}  \right) - \\
 - \frac{{\omega ^{2}}}{{c^{2}}}\,\frac{{1}}{{2\pi
}}\sum\limits_{m'n'\lambda '} {\int {\vec {Y}^{\lambda ^{\ast} }_{mn} \left(
{\vec {r}_{ \bot}  ,k_{z}}  \right)\hat {\chi} \left( {\vec {r}} \right)\vec
{Y}^{\lambda '}_{m'n'} \left( {\vec {r}_{ \bot}  ,k'_{z}}  \right)e^{ -
i\left( {k_{z} - k'_{z}}  \right)z}} \,} d^{2}r_{ \bot}  C_{m'n'}^{\lambda
'} \left( {k'_{z} ,\omega}  \right)dk'_{z} dz = \\
 = - \frac{{\omega _{L}^{2} \left( {k_{mn}^{2} c^{2} - \omega ^{2}}
\right)}}{{\gamma c^{4}\left( {\omega - \vec {k}_{mn} \vec {u}}
\right)^{2}}}\left\{ {\frac{{1}}{{2\pi} }\left| {\int {\vec {u}\,\vec
{Y}^{\lambda} _{mn} ( {\vec {k}_{ \bot}  ,k_{z}}  )d^{2}k_{ \bot
}} }  \right|^{2}} \right\}C^{\lambda} _{mn} \left( {k_{z} ,\omega}
\right), \\
 \end{array}
\end{equation}

\noindent where $\vec {Y}^{\lambda} _{mn} ( {\vec {k}_{ \bot}
,k_{z}}  ) = \int {e^{ - i\vec {k}_{ \bot}  \vec {r}_{ \bot}
} \vec {Y}_{mn}^{\lambda} } \left( {\vec {r}_{ \bot}  ,k_{z}}
\right)d^{2}r_{ \bot}  $.

{The system of equations (\ref{eq10}) in the approximation linear
over perturbation describes the electromagnetic field modes, which
are induced by an electron beam in the finite volume of a
rectangular waveguide due to the periodic disturbance produced by
a photonic crystal.}

\section{Radiative instability of a relativistic electron beam moving
in a bounded photonic crystal}

{The above obtained system of equations (\ref{eq10}) enables to
derive the dispersion equation for a bounded photonic crystal and
to analyze conditions, when electron beam instability presents. }
Let us consider the sums in the left-hand side of equation
(\ref{eq10}):

\begin{equation}
\label{eq10.1}
\begin{array}{l}
\sum\limits_{m'n'\lambda '} {\int {\vec {Y}^{\lambda ^{\ast} }_{mn} \left(
    {\vec {r}_{ \bot}  ,k_{z}}  \right)\hat {\chi} \left( {\vec {r}} \right)\vec
    {Y}^{\lambda '}_{m'n'} \left( {\vec {r}_{ \bot}  ,k'_{z}}  \right)e^{ -
        i\left( {k_{z} - k'_{z}}  \right)z}} \,} d^{2}r_{ \bot}  C_{m'n'}^{\lambda
'} \left( {k'_{z} ,\omega}  \right)dk'_{z} dz = \\
\sum\limits_{m'n'\lambda '}{\int C_{m'n'}^{\lambda
        '} \left( {k'_{z} ,\omega}\right)  {\int {\vec {Y}^{\lambda ^{\ast} }_{mn} \left(
        {\vec {r}_{ \bot}  ,k_{z}}  \right)\hat {\chi} \left( {\vec {r}} \right)\vec
        {Y}^{\lambda '}_{m'n'} \left( {\vec {r}_{ \bot}  ,k'_{z}}  \right)e^{ -
            i\left( {k_{z} - k'_{z}}  \right)z}} \,} d^{2}r_{ \bot}  dz\, dk'_{z}}
        \end{array}
        \end{equation}
and analyze integrals
\begin{equation}
 {\int {\vec {Y}^{\lambda ^{\ast} }_{mn} \left(
            {\vec {r}_{ \bot}  ,k_{z}}  \right)\hat {\chi} \left( {\vec {r}} \right)\vec
            {Y}^{\lambda '}_{m'n'} \left( {\vec {r}_{ \bot}  ,k'_{z}}  \right)e^{ -
                i\left( {k_{z} - k'_{z}}  \right)z}} \,d^{2}r_{ \bot}
                dz}.
                \label{eq:I}
\end{equation}
to evaluate what terms mostly contribute to the considered sums.
Using (\ref{eq:chi}) and representing eigenfunctions $\vec
{Y}_{mn}^{\lambda} \left({\vec {r}_{ \bot} ,k_{z}}  \right)$ of a
rectangular waveguide \cite{15,16,17} by combinations  of sines
and cosines of the form $sin\frac{{\pi m}}{{a}}x$, $cos\frac{{\pi
m}}{{a}}x$, $sin\frac{{\pi n}}{{b}}y$, $cos\frac{{\pi n}}{{b}}y$
 (i.e., in fact, the combinations $e^{i\frac{{\pi
m}}{{a}}x},e^{i\frac{{\pi n}}{{b}}y}$) integrals (\ref{eq:I}) can
be transformed to the expressions of the form as follows:
\begin{eqnarray}
\label{eq:Ix} I_x & = & \int {e^{ - i\frac{{\pi m}}{{a}}x}}
\sum\limits_{i} {\hat {\chi} _{cell} } \left( {x - x_{i} ,y -
y_{i} ,z - z_{i}} \right)e^{i\frac{{\pi
        m'}}{{a}}x}dx,  \\
I_y & = & \int {e^{ - i\frac{{\pi n}}{{b}}y}} \sum\limits_{i}
{\hat {\chi} _{cell} } \left( {x - x_{i} ,y - y_{i} ,z - z_{i}}
\right)e^{i\frac{{\pi
        n'}}{{b}}y}dy
        \label{eq:Iy}
\end{eqnarray}
Substitution of variables $x - x_{i} = \eta_1 $ in (\ref{eq:Ix})
and $y - y_{i} = \eta_2$  in (\ref{eq:Iy}) produces the sums of
the form

\[
S_{x} = \sum\limits_{f_1=1}^{N_{x}} {e^{ - i\frac{{\pi}
}{{a}}\left( {m - m'} \right)d_{x} f_1} }, \, S_{y} =
\sum\limits_{f_2=1}^{N_{y}} {e^{ - i\frac{{\pi} }{{b}}\left( {n -
n'} \right)d_{y} {f_2}} },
\]

\noindent where $d_{x}$ and $d_{y}$ are the periods of the
photonic crystal along $x$ and $y$ axes, $N_{x}=\frac{a}{d_x} $
and $N_{y}=\frac{b}{d_y}$ are the number of cells along $x$ and
$y$ axes, respectively; coordinates of different cells $x_{i} =
d_{x} f_{1} $, $y_{i} = d_{y} f_{2}$ are determined by integers
$f_{1}$ and $f_2$.
To estimate, what values can $S_{x}$  take (the same reasoning is
valid for $S_{y}$), the above presented expression can be
rewritten as follows:
\begin{equation}
\label{eq11} S_{x} = \sum\limits_{f_1=1}^{N_{x}} {e^{ -
i\frac{{\pi} }{{a}}\left( {m - m'} \right)d_{x} f_1} } =
e^{i\frac{{\pi} }{{2a}}\left( {m - m'} \right)\left( {N_{x} - 1}
\right)d_{x}} \frac{{sin  \frac{{\pi \left( {m - m'} \right)d_{x}
N_{x} }}{{2a}} } }{{sin\frac{{\pi \left( {m - m'} \right)d_{x}}
}{{2a}}}}.
\end{equation}
Using (\ref{eq11}) for $m - m' = 0$ one can obtain $S_{x} =
N_{x}$.
When $m - m' = 1$, a simple reasoning enables to get what $S_{x}$
is equal to: factor
 $d_{x} N_{x} = a$ and
 hence, the numerator {is equal to} $sin\frac{{\pi} }{{2}} = 1$, while in the denominator $sin\frac{{\pi
d_{x}} }{{2a}} \approx \frac{{\pi} }{{2N_{x}} }$.
Therefore, the ratio $\frac{{S_{x} \left( {m - m' = 1}
\right)}}{{S_{x} \left( {m - m' = 0} \right)}} = \frac{{2}}{{\pi}
} \approx 0.6$.
With growing difference $(m - m')$, the contribution to the sum of
the next terms diminishes provided the following equality is
fulfilled
\begin{equation}
\label{eq12}
\frac{{\pi \left( {m - m'} \right)d_{x}} }{{2a}} = \pi {\rm P},
\end{equation}
\noindent where ${\rm P} = \pm 1, \pm 2...$; in these cases the
sum $S_{x} = N_{x} $.

Fulfillment of conditions declared by equalities like
 (\ref{eq12}) is equivalent to fulfillment of conditions
$k_{xm} - k'_{xm'} = \tau _{x} $  (i.e., $k'_{xm'} = k_{xm} - \tau
_{x} $) and $k_{yn} - k'_{yn'} = \tau _{y} $ (i.e., $k'_{yn'} =
k_{yn} - \tau _{y} $), where $\tau _{x} = \frac{{2\pi} }{{d_{x}}
}F$ and $\tau _{y} = \frac{{2\pi} }{{d_{y}} }F'$ are $x$- and
$y$-components of the reciprocal lattice vector of the photonic
crystal, respectively,  $F,F' = 0, \pm 1, \pm 2...$ .
Therefore, the major contribution to the sums in the left-hand
side of equation (\ref{eq10}) comes from the amplitudes
$C_{m'n'}^{\lambda '} \left( {k'_{z} ,\omega} \right) \equiv
C^{\lambda '}\left( {\vec {k}_{ \bot mn} - \vec {\tau} _{ \bot}
,k_{z} - \tau _{z} ,\omega} \right) = C^{\lambda '}\left( {\vec
{k}_{mn} - \vec {\tau} ,\omega} \right)$.

{Hereinafter, when describing electron beam instability, we
consider only modes those satisfy the equalities similar to
(\ref{eq12}). The contribution of other modes is supposed to be
suppressed.}
Thus, system of equations (\ref{eq10}) reads as follows:

\begin{equation}
\label{eq13}
\begin{array}{l}
 \left( {\vec {k}_{mn}^{2} - \frac{{\omega ^{2}}}{{c^{2}}}}
\right)C^{\lambda} \left( {\vec {k}_{mn} ,\omega}  \right) - \frac{{\omega
^{2}}}{{c^{2}}}\sum\limits_{\lambda '\tau}  {\chi _{mn}^{\lambda \lambda '}
\left( {\vec {\tau} } \right)} C^{\lambda '}\left( {\vec {k}_{mn} - \vec
{\tau} ,\omega}  \right) = \\
 - \frac{{\omega _{L}^{2} \left( {k_{mn}^{2} c^{2} - \omega ^{2}}
\right)}}{{\gamma c^{4}\left( {\omega - \vec {k}_{mn} \vec {u}}
\right)^{2}}}\left\{ {\frac{{1}}{{2\pi} }\left| {\int {\vec {u}\,\vec
{Y}^{\lambda} _{mn} \left( {\vec {k}_{ \bot}  ,k_{z}}  \right)d^{2}k_{ \bot
}} }  \right|^{2}} \right\}C^{\lambda} \left( {\vec {k}_{mn} ,\omega}
\right), \\
 \end{array}
\end{equation}

\noindent
i.e.,

\begin{equation}
\label{eq14}
\begin{array}{l}
 \left( {\vec {k}_{mn}^{2} - \frac{{\omega ^{2}}}{{c^{2}}}\left[ {1 + \chi
_{mn}^{\lambda \lambda}  \left( {0} \right) - \frac{{\omega
_{L}^{2} \left( {k_{mn}^{2} c^{2} - \omega ^{2}} \right)}}{{\omega
^{2}\gamma c^{2}\left( {\omega - \vec {k}_{mn} \vec {u}}
\right)^{2}}}\left\{ {\frac{{1}}{{2\pi }}\left| {\int {\vec
{u}\,\vec {Y}^{\lambda} _{mn} \left( {\vec {k}_{ \bot} ,k_{z}}
\right)d^{2}k_{ \bot} } }  \right|^{2}} \right\}} \right]}
\right)C^{\lambda} \left( {\vec {k}_{mn} ,\omega}  \right) \\
 - \frac{{\omega ^{2}}}{{c^{2}}}\sum\limits_{\lambda '\tau}  {\chi
_{mn}^{\lambda \lambda '} \left( {\vec {\tau} } \right)} C^{\lambda '}\left(
{\vec {k} - \vec {\tau} ,\omega}  \right) = 0 \\
 \end{array}
\end{equation}

\noindent where $\chi _{mn}^{\lambda \lambda '} \left( {\tau}
\right) = \frac{{1}}{{d_{z}} }\int {\vec {Y}^{\lambda \ast} _{mn}
\left( {\vec {r}_{ \bot}  ,k_{z}}  \right)\hat {\chi} \left( {\vec
{r}_{ \bot}  ,\tau _{z}} \right)} \vec {Y}_{m'n'}^{\lambda '}
\left( {\vec {r}_{ \bot}  ,k_{z} - \tau _{z}}  \right)d^{2}r_{
\bot}  ~$, $\hat {\chi} \left( {\vec {r}_{ \bot} ,\tau _{z}}
\right) = \sum\limits_{x_{i} ,y_{i}}  {\int {\hat {\chi }_{cell}
\left( {x - x_{i} ,y - y_{i} ,\zeta}  \right)}}  \,e^{ - i\tau
_{z} \zeta} d\zeta $, $m'$ and  $n'$ are determined by conditions
like (\ref{eq12}),
{$\omega _{L} $ is the Langmuir frequency, $\omega _{_{L}} ^{2} =
\frac{{4\pi e^{2}\rho _{b0} }}{{m}}$.}

Since this system of equations is homogeneous, for the existence
of a nontrivial solution the system determinant must vanish. This
condition determines the dispersion equation.

Note that expression $\vec{k}_{mn} - \vec{k} = \vec{\tau} $ is
very much similar to Bragg condition and even converts to it, when
the transversal dimensions of a waveguide $a$ and $b$ tend to
infinity. Actually, when $a,\,b \rightarrow \infty$ the spectrum
of eigenvalues becomes continuous and the set of wave vectors
$\vec {k}_{mn}$ converts to wave vector $\vec {k}^{\prime}$ of the
wave propagating in the direction determined by Bragg condition
$\vec {k}^{\prime}-\vec{k}=\vec{\tau}$, which in more familiar
notation reads as
\begin{equation}
2 d_g \sin \theta_B= m \lambda_B, \label{Bragg}
\end{equation}
where $d_g$ is the diffraction grating period, $\theta_B$ is the
Bragg angle, which determines the direction of diffracted wave
propagation, $\lambda_B$ is the wavelength of the diffracted wave,
$m$ is an integer number.
Smooth rotation of diffraction grating with respect to electron
velocity at angle $\varphi$ converts $d_g$ in (\ref{Bragg}) to
$d_g \cos \varphi$, thus making $\lambda_B$ and $\theta_B$  in
(\ref{Bragg}) slowly changing in a wide range.

System of equations (\ref{eq14}) is similar to that describing
instability of a beam passing through an infinite crystal
\cite{18,19}. However, the coefficients appearing in these two
systems are differently defined: for an infinite crystal, the wave
vectors have continuous spectrum of eigenvalues rather than the
discrete spectrum relevant for a bounded photonic crystal.
These equations enable one to define dependence $k\left( {\omega}
\right)$ for the waves propagating in the crystal.
Matching the incident wave packet with the set of waves
propagating inside the photonic crystal by the boundary
conditions, one can obtain the explicit solutions of the
considered equations:
the result obtained is formally analogous to that given in
\cite{20}.

According to (\ref{eq14}), the expression within the square
brackets acts as dielectric permittivity
{$\varepsilon $ of the crystal in case when diffraction can be
neglected:
\begin{equation}
\varepsilon = n^{2} = 1 + \chi _{mn}^{\lambda \lambda} \left( {0}
\right) - \frac{{\omega _{L}^{2} \left( {k_{mn}^{2} c^{2} - \omega
^{2}} \right)}}{{\omega ^{2}\gamma c^{2}\left( {\omega - \vec
{k}_{mn} \vec {u}} \right)^{2}}}\left\{ {\frac{{1}}{{2\pi} }\left|
{\int {\vec {u}\,\vec {Y}^{\lambda} _{mn} \left( {\vec {k}_{ \bot}
,k_{z}} \right)d^{2}k_{ \bot} } }  \right|^{2}} \right\},
\label{eq:epsilon0}
\end{equation}
 $n$  is the refractive index i.e. photonic crystal
 acts as a medium that can be described
  by a ceratin refractive index $n$ or the dielectric permittivity
  $\varepsilon$. The refractive index of
the photonic crystal in the absence of the beam $n_{0} $ is
determined by $n_{0}^{2} = \varepsilon _{0} = 1 + \chi
_{mn}^{\lambda \lambda} \left( {0} \right)$.
 }
Two terms contribute to  dielectric permittivity
(\ref{eq:epsilon0}): scattering of waves by the unit cell of the
crystal and scattering of waves by beam electrons. The latter is
given by the term proportional to $\omega _{L}^{2} $ and increases
when $\omega \to \vec {k}\vec {u}$.

\subsection{Electron beam instability in a bounded photonic crystal beyound conditions enabling Bragg's diffraction}\label{ssec1}

Let us first assume that {there are no waves for which Bragg's
diffraction conditions are fulfilled}. Then the amplitudes of
diffracted waves are assumed to be small. In this case the sum
over $\tau $ in (\ref{eq14}) can be dropped, and the conditions
for the wave existence are given by putting equal to zero {the
expression left to $C^{\lambda} \left( {\vec {k}_{mn} ,\omega}
\right)$.
This requirement can be written in the form

\[
\left( {\omega - k_{z} u} \right)^{2}\left( {k_{mn}^{2} - \frac{{\omega
^{2}}}{{c^{2}}}n_{0}^{2}}  \right) = - \frac{{\omega _{L}^{2} \left(
{k_{mn}^{2} c^{2} - \omega ^{2}} \right)}}{{\gamma c^{4}}}\left\{
{\frac{{1}}{{2\pi} }\left| {\int {\vec {u}\,\vec {Y}^{\lambda} _{mn} \left(
{\vec {k}_{ \bot}  ,k_{z}}  \right)d^{2}k_{ \bot} } }  \right|^{2}}
\right\},
\]

\noindent where velocity $\vec {u} || oz$,
therefore,

\begin{equation}
\label{eq15} \left( {k_{z}^{2} - \left( {\frac{{\omega
^{2}}}{{c^{2}}}n_{0}^{2} - \varkappa_{mn}^{2}}  \right)}
\right)\left( {\omega - k_{z} u} \right)^{2} = - \frac{{\omega
_{L}^{2} \left( {k_{mn}^{2} c^{2} - \omega ^{2}} \right)}}{{\gamma
c^{4}}}\left\{ {\frac{{1}}{{2\pi} }\left| {\int {\vec {u}\,\vec
{Y}^{\lambda} _{mn} \left( {\vec {k}_{ \bot}  ,k_{z}}
\right)d^{2}k_{ \bot} } }  \right|^{2}} \right\}
\end{equation}

Since the nonlinearity is insignificant, let us consider  the
spectrum of the waves of equation (\ref{eq15}) with
  zero right-hand side as zero-order
 approximation.

Let us concern with the case when $\omega - k_{z} u \to 0$ (i.e.,
the Cherenkov radiation condition is fulfilled) and $\left(
{k_{z}^{2} - \left( {\frac{{\omega ^{2}}}{{c^{2}}}n_{0}^{2} -
\varkappa _{mn}^{2}}  \right)} \right) \to 0$, i.e, the
electromagnetic wave can propagate in a photonic crystal without
the beam.
With zero right-hand side equation (\ref{eq15}) reads as follows:

\begin{equation}
\label{eq16} \left( {k_{z}^{2} - \left( {\frac{{\omega
^{2}}}{{c^{2}}}n_{0}^{2} - \varkappa _{mn}^{2}}  \right)} \right)
= 0, \quad \left( {\omega - k_{z} u} \right) = 0.
\end{equation}

The frequency of radiation produced by the electron beam is
determined by interception of dispersion curve $\omega(k_z)$ with
the beam line $\left( {\omega - k_{z} u} \right) = 0 $ (see
Fig.\ref{fig1}) that means synchronism of wave and beam electrons.
Interceptions can occur at different values of $v_{gr}=\frac{d
\omega}{d k}$: at $v_{gr}>0$ wave travels forward along beam
velocity (such situation is typical for travelling wave tubes);
wave with $v_{gr}<0$ moves backward enabling generation of
backward wave oscillators \cite{Benford}.

\begin{figure}[htbp]
    \centerline{\includegraphics[width=12 cm]{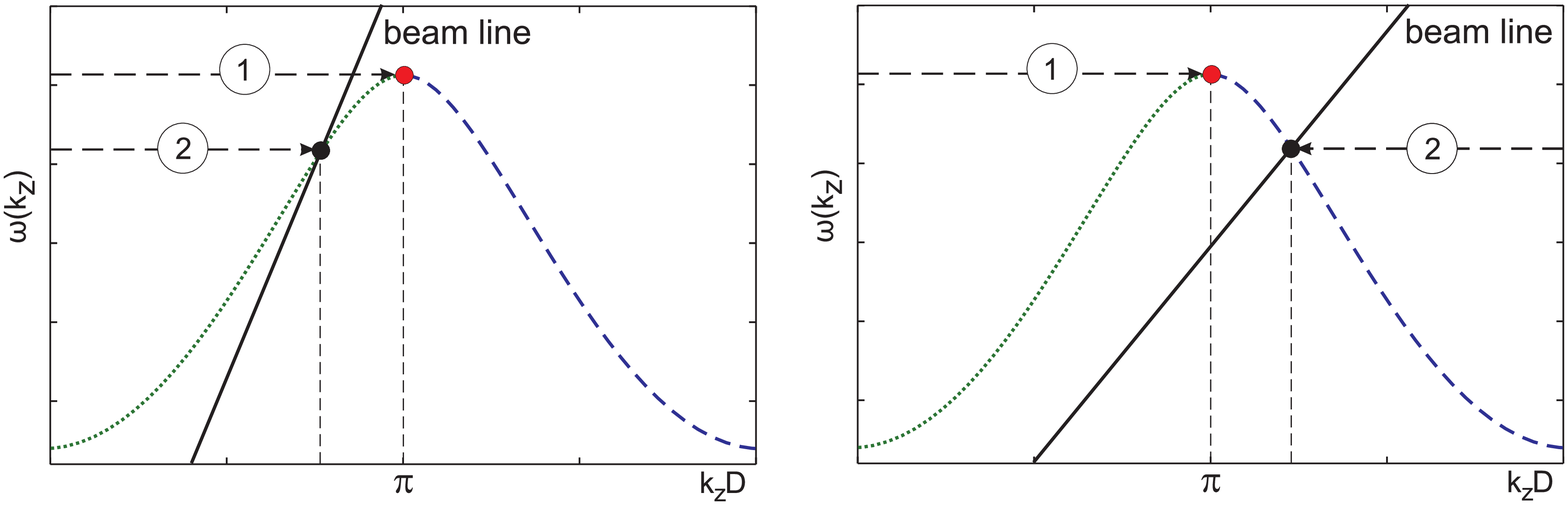}}
    \caption{Schematic drawing of dispersion curve $\omega(k_z)$. Point marked 1 corresponds to $v_{gr}=0$ ($\pi$-point), it matches $k_z D=\pi$,
    where $D$ is the period of periodic structure. Points marked 2 correspond to beam-wave synchronism points. Left plot is for TWT,
    right plot corresponds to BWO case}
    \label{fig1}
\end{figure}

The roots of equation (\ref{eq16}) are
\begin{equation}
\label{eq17} k_{1z} = {\pm}\frac{{\omega} }{{c}}\sqrt {n_{0}^{2} -
\frac{{\varkappa _{mn}^{2} c^{2}}}{{\omega ^{2}}}} , \quad k'_{1z}
= - k_{1z} , \quad k_{2z} = \frac{{\omega} }{{u}}.
\end{equation}

Since  $k_{2z} = \frac{{\omega} }{{u}}
> 0$ in view of the Cherenkov condition, we are concerned with propagation in the photonic crystal of the
wave with $k_{1z} > 0$ {(i.e. only sign ''+'' rests in
(\ref{eq17}))}. In this case in the equation for $k_{z} $, one can
take $\left( {k_{z} - k_{1z}} \right) \left( {k_{z} + k_{1z}}
\right) \approx 2k_{1z} \left( {k_{z} - k_{1z}} \right)$ and
rewrite equation (\ref{eq15}) as follows:

\begin{equation}
\label{eq18}
\left( {k_{z} - k_{1z}}  \right)\left( {k_{z} - k_{2z}}  \right)^{2} = -
\frac{{\omega _{L}^{2} \omega ^{2}\left( {n_{0}^{2} - 1} \right)}}{{2k_{1z}
u^{2}\gamma c^{4}}}\left\{ {\frac{{1}}{{2\pi} }\left| {\int {\vec {u}\,\vec
{Y}^{\lambda} _{mn} \left( {\vec {k}_{ \bot}  ,k_{z}}  \right)d^{2}k_{ \bot
}} }  \right|^{2}} \right\}
\end{equation}

\noindent
i.e.,

\begin{equation}
\label{eq19} \left( {k_{z} - k_{1z}}  \right)\left( {k_{z} -
k_{2z}}  \right)^{2} = - A,
\end{equation}

\begin{equation}
A=\frac{{\omega _{L}^{2} \omega ^{2}\left( {n_{0}^{2} - 1}
\right)}}{{2 {k_{1z}} u^{2}\gamma c^{4}}}\left\{
{\frac{{1}}{{2\pi} }\left| {\int {\vec {u}\,\vec {Y}^{\lambda}
_{mn} \left( {\vec {k}_{ \bot}  ,k_{z}} \right)d^{2}k_{ \bot }} }
\right|^{2}} \right\}, \label{eq:A}
\end{equation}

\noindent where $A$ is real and $A > 0$ (as to enable Cherenkov
effect, it is necessary to have $n_{0}^{2} > 1$).

Thus, $k_{z}$ satisfies cubic equation (\ref{eq19}). Roots $k_{1z}
$ and $k_{2z} $ coincide ($k_{1z} = k_{2z} $), when the particle
velocity satisfies condition as follows:

\begin{equation}
\label{eq20} u = \frac{{c}}{{\sqrt {n_{0}^{2} - \frac{{\varkappa
_{mn}^{2} c^{2}}}{{\omega ^{2}}}}} }.
\end{equation}

\noindent For $k_{1z} = k_{2z}$ substitution $\xi = k_z - k_{1z} $
in (\ref{eq19}) gives equation

\begin{equation}
\label{eq21} \xi ^{3} = - A,
\end{equation}

 \noindent which has three solutions
\begin{equation}
\xi ^{(1)} = - \sqrt[{3}]{{A}}, ~
 \xi
^{(2,3)} = \frac{{1}}{{2}}\left( {1 \pm i\sqrt {3}}  \right)
\sqrt[{3}]{{A}}. \label{eq:root3}
\end{equation}

{ The state corresponding to solution {$\xi ^{(3)} =
\frac{{1}}{{2}}\left( {1 - i\sqrt {3}}  \right)\sqrt[{3}]{{A}}$}
grows with $z$ growing that indicates the presence of instability
in a beam \cite{21,Miller}.}
In this case the beam instability increment
\begin{equation}
\delta_0^{(3)} \sim Im\,k_{z} =
{Im\,\xi ^{(3)}}
\sim\sqrt[{3}]{{\omega_L^2}} \sim \sqrt[{3}]{{\rho_{b0}} }
\label{eq:increment3}
\end{equation}
and the threshold current density (a minimum beam current density
required for oscillations to start spontaneously) is determined by
the well-known law $j_{thr} \sim \frac{1}{L^3}$ ( see equation
(11.7) in \cite{Granatstein} or equation (8.63) in
\cite{Tsimring}).

\subsection{Electron beam radiative instability in a bounded
photonic crystal in the vicinity of $\pi$-point}\label{ssec2}

Let us now suppose that in a bounded photonic crystal there is a
wave, for which Bragg's diffraction conditions are fulfilled. In
this case the roots of dispersion equation are close to each other
that happens, when $k_z$ is in the vicinity of $\pi$-point
\cite{23,24,25}.
Such assumptions mean that in (\ref{eq14})
the wave amplitude $C_{mn} \left( {\vec{k}_{mn} + \vec {\tau} }
\right)$ is comparable with the amplitude $C_{mn} \left( {\vec
{k}_{mn}} \right)$. {This is in contrast to assumptions used to
derive (\ref{eq18}) and, for this case conclusions obtained in
subsection \ref{ssec1} are not valid.}
We also assume that $\chi \ll 1$, thus only the equations for
these two amplitudes remain in (\ref{eq14}) similar to the
standard diffraction theory for an infinite crystal \cite{23, 24}.

{When $\chi \sim 1$ one should consider more waves (see for
example \cite{Yablonovitch2}) and the number of equations in
(\ref{eq14}) would be greater.}

Analysis of diffraction of  wave $\lambda $, which electric vector
is parallel to the plane $\left( {y,z} \right)$ (a TM-wave) gives:

\begin{equation}
\label{eq22} \left[ {k_{mn}^{2} - \frac{{\omega
^{2}}}{{c^{2}}}\varepsilon} \right]C^{\lambda} \left( {\vec
{k}_{mn} ,\omega}  \right) - \frac{{\omega ^{2}}}{{c^{2}}}\chi
_{mn}^{\lambda \lambda}  \left( { - \vec {\tau} }
\right)C^{\lambda} \left( {\vec {k}_{mn} + \vec {\tau} ,\omega}
\right) = 0,
\end{equation}

\[
\left[ {\left( {\vec {k}_{mn} + \vec {\tau} } \right) - \frac{{\omega
^{2}}}{{c^{2}}}\varepsilon _{0}}  \right]C^{\lambda} \left( {\vec {k}_{mn} +
\vec {\tau} ,\omega}  \right) - \frac{{\omega ^{2}}}{{c^{2}}}\chi
_{mn}^{\lambda \lambda}  \left( {\vec {\tau} } \right)C^{\lambda} \left(
{\vec {k}_{mn} ,\omega}  \right) = 0.
\]

Since the term containing $( {\omega - ( {\vec {k} + \vec {\tau} }
)\vec {u}} )^{ - 1}$ is small when $( {\omega - \vec {k}\vec {u}}
)$ vanishes, in the second equation it is dropped.

The dispersion equation defining the relation between $k_{z} $ and
$\omega $ is obtained by equating to zero the determinant of the
system (\ref{eq22}) and has a form:

\begin{equation}
\label{eq23}
\begin{array}{l}
 \left[ {\left( {k_{mn}^{2} - \frac{{\omega ^{2}}}{{c^{2}}}\varepsilon _{0}
} \right)\left( {\left( {\vec {k}_{mn} + \vec {\tau} } \right)^{2} -
\frac{{\omega ^{2}}}{{c^{2}}}\varepsilon _{0}}  \right) - \frac{{\omega
^{4}}}{{c^{4}}}\chi _{\tau}  \chi _{ - \tau} }  \right]\left( {\omega -
k_{z} u} \right)^{2} = \\
 - \frac{{\omega _{L}^{2}} }{{\gamma c^{4}}}\left\{ {\frac{{1}}{{2\pi
}}\left| {\int {\vec {u}\,\vec {Y}^{\lambda} _{mn} \left( {\vec
{k}_{ \bot} ,k_{z}}  \right)d^{2}k_{ \bot} } }  \right|^{2}}
\right\}\left( {k_{mn}^{2} c^{2} - \omega ^{2}} \right)\left(
{\left( {\vec {k}_{mn} + \vec {\tau} } \right)^{2} - \frac{{\omega
^{2}}}{{c^{2}}}\varepsilon _{0}}  \right).
 \end{array}
\end{equation}

The right-hand side of equation (\ref{eq23}) is small, therefore,
one can again seek the solution near the points, where the
right-hand side is zero, that means fulfillment of Cherenkov
radiation condition
 and excitation of the wave, which can
propagate in the waveguide:

\begin{equation}
\label{eq24}
\begin{array}{l}
 \left( {k_{z}^{2} - \left( {\frac{{\omega ^{2}}}{{c^{2}}}\varepsilon _{0} -
\varkappa _{mn}^{2}}  \right)} \right)\left( {\left( {k_{z} +
\tau} \right)^{2} - \left( {\frac{{\omega
^{2}}}{{c^{2}}}\varepsilon _{0} - \left( {\vec {\varkappa} _{mn} +
\vec {\tau} _{ \bot} }  \right)^{2}} \right)} \right)
- \frac{{\omega ^{4}}}{{c^{4}}}\chi _{\tau}  \chi _{ - \tau}  = 0 \\
 \left( {k_{z} - \frac{{\omega} }{{u}}} \right)^{2} = 0 \\
 \end{array}
\end{equation}

The roots for system of equations (\ref{eq24}) are sought {in the
vicinity of condition $k_{mn}^{2} \approx \left( {\vec {k}_{mn} +
\vec {\tau} } \right)^2$}, by substitution $\xi =k_{z}- k_{z0}$,
which gives the expressions as follows:

\begin{equation}
\label{eq25} k_{z} = k_{z0} + \xi , \quad k_{z}^{2} = k_{z0}^{2} +
2k_{z0} \xi + \xi ^{2}, \quad k_{z0}^{2} = \frac{{\omega
^{2}}}{{c^{2}}}\varepsilon _{0} - \varkappa _{mn}^{2} ; \quad
k_{z0} = \frac{{\omega} }{{c}}\sqrt {\varepsilon _{0} -
\frac{{\varkappa _{mn}^{2} c^{2}}}{{\omega ^{2}}}}
\end{equation}

\[
\left( {k_{z} + \tau _{z}}  \right)^{2} = \left[ {\left( {k_{z0} +
\tau _{z} } \right) + \xi}  \right]^{2} = \left( {k_{z0} + \tau
_{z}}  \right)^{2} + 2\left( {k_{z0} + \tau _{z}}  \right)\xi +
\xi ^{2}
\]

Hence, transformation of expressions as follows

\begin{equation}
\label{eq26}
\begin{array}{l}
 \left( {k_{z0} + \tau _{z}}  \right)^{2} + \left( {\vec {\varkappa} _{mn} +
\vec {\tau} _{ \bot} }  \right)^{2} + 2\left( {k_{z0} + \tau _{z}}
\right)
+ 2\left( {k_{z0} + \tau _{z}}  \right)\xi + \xi ^{2} = \\
 \left( {\vec {k}_{mn} + \vec {\tau} } \right)^{2} + 2\left( {k_{z0} + \tau
_{z}}  \right)\xi + \xi ^{2} = k_{0mn}^{2} + 2\vec {k}_{0mn} \vec
{\tau} + \tau ^{2} + 2\left( {k_{z0} + \tau _{z}}  \right)\xi +
\xi ^{2}.
 \end{array}
\end{equation}

enables to render the first equation in (\ref{eq24}) as follows
\[
2k_{z0} \xi \left( {2\left( {k_{z0} + \tau _{z}}  \right)\xi +
\left( {2\vec {k}_{0mn} \vec {\tau}  + \tau ^{2}} \right)} \right)
- \frac{{\omega ^{4}}}{{c^{4}}}\chi _{\tau}  \chi _{ - \tau}  = 0,
\]
which is equivalent to
\begin{equation}
 4k_{z0} \left( {k_{z0} + \tau _{z}}  \right)\xi ^{2}
+ 2k_{z0} \left( {2\vec {k}_{0mn} \vec {\tau}  + \tau ^{2}}
\right)\xi - \frac{{\omega ^{4}}}{{c^{4}}}\chi _{\tau}  \chi _{ -
\tau}  = 0. \nonumber
\end{equation}
Thus, the second-order equation
\begin{equation}
 \xi ^{2} + \frac{{\left( {2\vec {k}_{0mn} \vec
{\tau} + \tau ^{2}} \right)}}{{\left( {k_{z0} + \tau _{z}}
\right)}}\xi - \frac{{\omega ^{4}}}{{c^{4}}}\frac{{\chi _{\tau}
\chi _{ - \tau} } }{{4k_{z0} \left( {k_{z0} + \tau _{z}} \right)}}
= 0. \label{eq27a}
\end{equation}
enables to get the following solutions for $\xi$:
\begin{equation}
\xi _{1,2} = - \frac{{\left( {2\vec {k}_{0} \vec {\tau}  + \tau
^{2}} \right)}}{{4\left( {k_{z0} + \tau _{z}}  \right)}} \pm \sqrt
{\left( {\frac{{2\vec {k}_{0} \vec {\tau}  + \tau ^{2}}}{{4\left(
{k_{z0} + \tau _{z}}  \right)}}} \right)^{2} + \frac{{\omega
^{4}}}{{c^{4}}}\frac{{\chi _{\tau}  \chi _{ - \tau} } }{{4k_{z0}
\left( {k_{z0} + \tau _{z}} \right)}}}. \label{xi_1,2}
\end{equation}

When $\left( {k_{z0} + \tau _{z}}  \right) = - \left| {k_{z0} +
\tau _{z}} \right|$, the second term in (\ref{xi_1,2}) could
become equal to zero. At the same time, the second equation in
(\ref{eq24}) must hold:
\[
\omega - k_{z} u = \omega - k_{z0} u - \xi u = 0.
\]
Consequently,

\begin{equation}
\xi_3 = \frac{{\omega - k_{z0} u}}{{u}} = \frac{{\omega} }{{u}} -
k_{z0} = \frac{{\omega} }{{u}} - \frac{{\omega} }{{c}}\sqrt
{\varepsilon _{0} - \frac{{\varkappa _{mn}^{2} c^{2}}}{{\omega
^{2}}}}= \frac{{\omega} }{{u}}\left( {1 - \beta \sqrt {\varepsilon
_{0} - \frac{{\varkappa _{mn}^{2} c^{2}}}{{\omega ^{2}}}}}
\right). \label{eq:xi_3}
\end{equation}
If $\varepsilon _{0} < 1$, then $\xi_3= \frac{{\omega} }{{u}} -
k_{z0}   > 0$,
Let solutions  $\xi _{1} $ and $\xi _{2} $ coincide ($\xi _{1} =
\xi _{2} $). This is possible at point

\[
\frac{{2\vec {k}_{0} \vec {\tau}  + \tau ^{2}}}{{4\left( {k_{z0} +
\tau _{z} } \right)}} = \pm \frac{{\omega
^{2}}}{{c^{2}}}\frac{{\sqrt {\chi _{\tau} \chi _{ - \tau} } }
}{{\sqrt {4k_{z0} \left| {k_{z0} + \tau _{z}}  \right|} }},
\]

\noindent here $k_{z0} + \tau _{z} < 0$,
and the following equality is fulfilled

 \[\frac{{\omega} }{{u}} - k_{z0} = \mp \frac{{\omega
^{2}}}{{c^{2}}}\frac{{\sqrt {\chi _{\tau}  \chi _{ - \tau} } }
}{{\sqrt {4k_{z0} \left| {k_{z0} + \tau _{z}}  \right|}} },
\]
 i.e.,
 \begin{equation}
 \frac{{\omega} }{{u}}
= k_{z0} \mp \frac{{\omega ^{2}}}{{c^{2}}}\frac{{\sqrt {\chi
_{\tau}  \chi _{ - \tau} } } }{{\sqrt {4k_{z0} \left| {k_{z0} +
\tau _{z}}  \right|}} } , \label{eq:root}
\end{equation}
where $k_{z0} = \frac{{\omega} }{{c}}\sqrt {\varepsilon _{0} -
\frac{{\varkappa _{mn}^{2} c^{2}}}{{\omega ^{2}}}}$. When
$\varepsilon _{0} < 1$ ratio $\frac{{\omega} }{{u}}
> k_{z0} $ (since $u < c$),  and for solution $ \frac{{\omega} }{{u}}
= k_{z0} - \frac{{\omega ^{2}}}{{c^{2}}}\frac{{\sqrt {\chi _{\tau}
\chi _{ - \tau} } } }{{\sqrt {4k_{z0} \left| {k_{z0} + \tau _{z}}
\right|}} }$ (\ref{eq:root}) the Cherenkov condition is not
fulfilled.

Now let us consider the solution $\frac{{\omega} }{{u}} = k_{z0} +
\frac{{\omega ^{2}}}{{c^{2}}}\frac{{\sqrt {\chi _{\tau}  \chi _{ -
\tau} } } }{{\sqrt {4k_{z0} \left| {k_{z0} + \tau _{z}}  \right|}}
}$. At $\tau _{z} < 0$ the difference $k_{z0} + \tau _{z} $ can be
reduced to make the sum on the right-side equal to $\frac{{\omega}
}{{u}}$, thus providing equality of all four dispersion equation
roots.

Note that for backward diffraction, which is conventional for
frequently used one-dimensional generators with a corrugated metal
waveguide (traveling-wave tube, backward-wave oscillator), such
coincidence of roots is impossible.
Indeed, suppose the solutions $\xi _{1} $ and $\xi _{2} $ coincide
in case of backward Bragg diffraction ($\left| {\tau _{z}} \right|
\approx 2k_{z0} $, $\tau _{z} < 0$). Then by substituting the
expressions for
\[
k_{z0} = \frac{{\omega} }{{c}}\sqrt {\varepsilon _{0} -
\frac{{\varkappa _{mn}^{2} c^{2}}}{{\omega ^{2}}}} \text{ and }
\varepsilon _{0} = n_{0}^{2} = 1 + \chi _{mn}^{\lambda \lambda}
\left( {0} \right)
\]
and retaining the first-order infinitesimal terms, the relation
\[\frac{{\omega} }{{u}} \approx k_{z0} + \frac{{\omega
^{2}}}{{c^{2}}}\frac{{\left| {\chi _{\tau} } \right|}}{{2k_{z0}}
}\] can be reduced to the form
\[
\frac{{\omega} }{{u}} \approx \frac{{\omega }}{{c}}\left( {1 -
\frac{{\left| {\chi _{mn}^{\lambda \lambda}  \left( {0} \right)}
\right|}}{{2}} - \frac{{\varkappa _{mn}^{2} c^{2}}}{{2\omega
^{2}}} + \frac{{\omega} }{{c}}\frac{{\left| {\chi _{\tau} }
\right|}}{{2}}} \right) < \frac{{\omega} }{{u}},
\]
i.e., the equality does not hold and the four-fold degeneracy is
impossible. Only the case of three-fold degeneration considered in
subsection \ref{ssec1} is possible.
{However, if $\varepsilon_{0}>1$ and is appreciably large, then in
a one-dimensional case, the four-fold degeneracy of roots is also
possible in a finite photonic crystal
%
}

Thus, the left-side of equation  (\ref{eq23}) has four solutions
($\xi_{1}$, $\xi_{2}$, and a double degenerated $\xi_{3}$). Hence,
equation (\ref{eq23}) can be written as follows:

\begin{equation}
\left( {\xi - \xi _{1}}  \right)\left( {\xi - \xi _{2}}  \right)
\left( {\xi - \xi _{3}}  \right)^{2} = -B, \label{eq:B}
\end{equation}
where $\xi_{1,2}$ and $\xi_{3}$ are defined by (\ref{xi_1,2}) and
(\ref{eq:xi_3}), respectively, $B$ is real, $B>0$,

\begin{equation}
B= \frac{{\omega _{L}^{2}} }{{4 k_{z0} (k_{z0}+\tau_z) u^2\gamma
c^{4}}}\left\{ {\frac{{1}}{{2\pi }}\left| {\int {\vec {u}\,\vec
{Y}^{\lambda} _{mn} \left( {\vec {k}_{ \bot} ,k_{z}}
\right)d^{2}k_{ \bot} } } \right|^{2}} \right\}\left( {k_{mn}^{2}
c^{2} - \omega ^{2}} \right)\left( {\left( {\vec {k}_{mn} + \vec
{\tau} } \right)^{2} - \frac{{\omega ^{2}}}{{c^{2}}}\varepsilon
_{0}}  \right). \label{eq:B_expression}
\end{equation}

\noindent If suppose $\xi _{1} = \xi _{2} = \xi _{3} $, then
equation (\ref{eq:B}) converts to
\begin{equation}
 \left( {\xi - \xi _{1}}  \right)^{4} = -B, \text{ i.e. }  \xi - \xi _{1}
 = \sqrt[{4}]{{-B}},
 \label{eq:4}
\end{equation}
 thus defining four solutions as follows:
 \begin{eqnarray}
\xi _{1}^{(1)} = \frac{1}{\sqrt{2}}(1 + \imath ) \sqrt[{4}]{{B}},~
\xi _{1}^{(2)} = \frac{1}{\sqrt{2}}(-1 + \imath )
\sqrt[{4}]{{B}},~
\xi_{1}^{(3)} =- \frac{1}{\sqrt{2}}( 1 + \imath )
\sqrt[{4}]{{B}},~
\xi_{1}^{(4)} =- \frac{1}{\sqrt{2}}(- 1 + \imath )
\sqrt[{4}]{{B}}.
\end{eqnarray}
Degeneration of roots of dispersion equation puts the synchronism
point to $\pi$-point of dispersion curve (see Fig. \ref{fig2}). In
this point amplitudes of direct and diffracted waves are
comparable to each other.

\begin{figure}[htbp]
    \centerline{\includegraphics[width=8 cm]{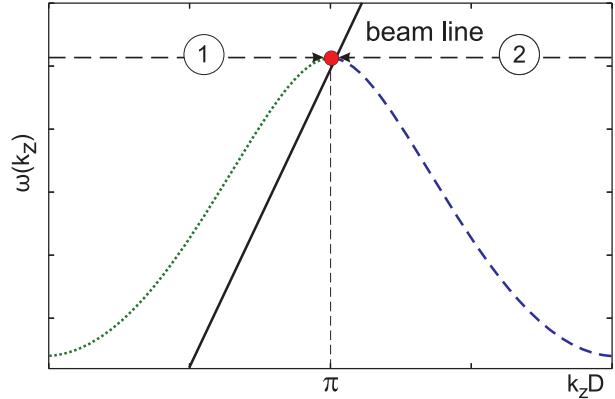}}
    \caption{Schematic drawing of dispersion curve $\omega(k_z)$ in case, when roots of dispersion equation are equal. }
    \label{fig2}
\end{figure}
Solutions $\xi_1^{(3)}$ and $\xi_1^{(4)}$  have negative imaginary
parts and cause exponential growth of field amplitude, thus they
are responsible for beam instability.
Increment of beam instability in case of four roots degeneracy:
\begin{equation}
\delta_0^{(4)} \sim Im\, k_z = Im \, \xi_1^{(3,4)} \sim
\sqrt[4]{\omega_L} \sim \sqrt[4]{\rho_{b0}}. \label{eq:root4}
\end{equation}
To compare increments  $\delta_0^{(3)}$ and $\delta_0^{(4)}$ let
us consider case when solutions $\xi_1 \ne \xi _{2}$ and $\xi_1
-\xi _{2} \gg \chi_{\tau}$ and study (\ref{eq:B}) for $\xi
\rightarrow \xi_1$,
thus reducing the order of equation:
\begin{equation}
\left( {\xi - \xi _{1}}  \right)  \left( {\xi - \xi _{3}}
\right)^{2} = -\frac{B}{{\xi_1 - \xi _{2}}}. \label{eq:Breduced}
\end{equation}
The obtained equation is conformable to (\ref{eq19}). Hence, the
ratio $\frac{\delta_0^{(4)}}{\delta_0^{(3)}}$ can be expressed as:
\begin{equation}
\frac{\delta_0^{(4)}}{\delta_0^{(3)}}=\sqrt{\frac{2}{3}}\sqrt[3]{\xi_1
- \xi _{2}} \frac{1}{\sqrt[12]{B}} \label{eq:ratio}
\end{equation}

For the sake of evaluations (\ref{eq:B_expression}) can be
replaced by the approximate expression for $B$  as follows:

\begin{equation}
B  \approx \frac{\omega_L^2}{\omega^2 \gamma} k_0^4 \chi_0^4,
\nonumber
 \label{eq:B_approx}
\end{equation}
hence, evaluation for $\delta_0^{(4)}$ and $\delta_0^{(3)}$ read:
\begin{eqnarray}
\delta_0^{(4)} \approx \frac{1}{\sqrt{2}}\sqrt[4]{k_0^4 \chi_0^4
\frac{\omega_L^2}{\omega^2 \,\gamma}} = \frac{1}{\sqrt{2}}
\frac{k_0 \chi_0}{\sqrt[4]{\gamma}}
\sqrt{\frac{\omega_L}{\omega}},\nonumber \\
\delta_0^{(3)} \approx \frac{\sqrt{3}}{2} \sqrt[3]{k_0^3 \chi_0^3
\frac{\omega_L^2}{\omega^2 \, \gamma}} = \frac{\sqrt{3}}{\sqrt{2}}
\,\, \delta_0^{(4)} \gamma^{-\frac{1}{12}}
\sqrt[6]{\frac{\omega_L}{\omega}} \nonumber
\end{eqnarray}
and ratio $\frac{\delta_0^{(4)}}{\delta_0^{(3)}}$ can be evaluated
by
\begin{equation}
\frac{\delta_0^{(4)}}{\delta_0^{(3)}}=\sqrt{\frac{2}{3}} \, \,
\gamma^{\frac{1}{12}}\sqrt[6] {\frac{\omega}{\omega_L}}.
\label{eq:ratio2}
\end{equation}

For terahertz range even for a beam with current density $j \sim
10^8$ A/cm$^2$ ($\omega_L \sim 10^8$\,Hz) expression
(\ref{eq:ratio2}) gives $\frac{\delta_0^{(4)}}{\delta_0^{(3)}} \gg
1$.

The typical length, at which instability develops, is inverse to
the increment value, therefore, along with (\ref{eq:ratio2}) the
following relations are also valid:

\begin{equation}
\frac{L^{(3)}}{L^{(4)}}=\frac{\delta_0^{(4)}}{\delta_0^{(3)}}=
\sqrt{\frac{2}{3}} \, \,\gamma^{\frac{1}{12}}\sqrt[6]
{\frac{\omega}{\omega_L}}. \label{eq:ratio3}
\end{equation}
\begin{equation}
L^{(4)} = \frac{1}{\delta_0^{(4)}} =\sqrt{2}
\frac{\sqrt[4]{\gamma}}{k_0 \chi_0}
\sqrt{\frac{\omega}{\omega_L}}\,, ~~L^{(3)} =
\frac{1}{\delta_0^{(3)}}=
\frac{2}{\sqrt{3}}\frac{\sqrt[3]{\gamma}}{k_0 \chi_0}
\sqrt[3]{\frac{\omega^2}{\omega_L^2}} \,, \nonumber
\label{eq:L_typical}
\end{equation}
providing evaluation for lengths ratio, at which instability
develops, $\frac{L^{(3)}}{L^{(4)}} \gg 1$.
Thus, generation in the vicinity of $\pi$-point (or, equally, in
case, when amplitudes of direct and diffracted waves approximate
to each other) gives a advantage of shorter length, at which
instability develops.

Let us analyze what values $L^{(3)}$  and $L^{(4)}$ can possess at
typical parameters of modern accelerators and radiation frequency
1 terahertz ($\lambda=3 \cdot 10^{-2}~$cm).
Suppose electron beam energy 8 Mev, bunch transversal size
250\,$\mu$m\,x\,250\,$\mu$m, bunch charge 25\,pC \cite{KEK}.
{Let us also consider a photonic crystal formed by parallel
metallic threads, which are parallel to the waveguide boundary
$\left( {y,z} \right)$.}
According to \cite{12} for such a photonic crystal with threads of
$10^{-3}~$cm diameter spaced $6 \cdot 10^{-2}~$cm the
susceptibility value is $\chi_0 \approx 3 \cdot 10^{-1}$.
For selected parameters, according to (\ref{eq:L_typical}), length
$L^{(4)} \approx 70$\,cm, while $L^{(3)}$ is more than 10 times
larger ($L^{(3)}\approx 900$\,cm).
Increase of electron beam current density could make length
$L^{(4)}$ even smaller. The same could be provided by increase
susceptibility value, but for $\chi_0 >1$ detailed numerical
analysis is necessary \cite{Yablonovitch2}.

The threshold generation conditions, i.e., the values of the
electron current and other parameters of the beam, at which
radiation begins to exceed the losses, can be obtained by solving
the boundary-value problem similar to how it was made in
\cite{Ch_6}. For instance the expression for the generation
threshold under the conditions of two--wave diffraction in the
case of cold beam reads:

\begin{equation} \quad \frac{1}{4\gamma } \frac{4\pi e^{2} }{\omega
^{2} m} \, \, \frac{j_{thr} }{u} \, \, \left|\frac{1}{u} \int
\vec{u}{\kern 1pt} \vec{Y}_{mn}^{ \lambda } \left(\vec{k}_{\bot }
,k_{z} \right)d^{2} k_{\bot } \right|^{2} f(y)=16
\left(\frac{\gamma _{0} c}{\vec{u}\vec{n}} \right)^{3} \frac{\beta
_{1} }{k^{5} \chi _{\tau } ^{2} L\, ^{5} } \label{eq:threshold}
\end{equation}
where $\gamma $ is the beam Lorentz factor; $\vec{u}$ is the
unperturbed velocity vector of the beam particles; $\vec{n}$ is
the unit normal vector to the crystal surface (directed toward the
interior);
$L$ is the crystal thickness; $\chi _{0} $ and $\chi _{\tau } $
are the Fourier expansion coefficients of the crystal dielectric
susceptibility;
$\beta _{1} =\gamma _{1} /\gamma _{0} $ is the diffraction
asymmetry factor; $\beta _{1} =\frac{\gamma _{1} }{\gamma _{0} }
=\frac{\vec{n}(\vec{k}+\vec{ \tau })}{\vec{n}\vec{k}} $,
$\gamma _{0} $ and $\gamma _{1} $ are the cosines of the angles
between the normal vector $\vec{n}$ and the wave vectors of the
transmitted $\vec{k}$ and diffracted $\vec{k}+ \vec{\tau }$ waves,
respectively; the subscript $\bot $ denotes the projection of the
vector on the plane perpendicular to $\vec{u}$; $f(y)$ is the
spectral function depending on detuning from the synchronism
conditions defined in \cite{Ch_6}:
\[
\label{GEQ110} f(y)=\sin y\frac{(2y+\pi n)\sin y-y(y+\pi n)\cos
y}{y^{3} (y+\pi n)^{3} } ,
\]
where $y=\frac{kRe(\xi _{2} )L}{2} $ and $\xi _{2} $ is the root
of the dispersion equation in the absence of the electron beam.
Expression (\ref{eq:threshold}) provides for $j_{thr}$ the
dependance on the crystal length $L$ as follows: $j_{thr} \sim
\frac{\beta_1}{L^5}$, which is different from that cited in
subsection \ref{ssec1} and \cite{Granatstein,Tsimring}.

{The analysis \cite{9,Ch_6,Ch_7} shows that with increasing the
number of diffracted waves, the law established in \cite{Ch_1,7,8}
is still valid: the instability increment appears to be
proportional to $\rho_{b0} ^{\frac{{1}}{{s + 3}}}$, where $s$ is
the number of waves emerging through diffraction. As a result, the
abrupt decrease in the threshold generation current also remains
in this case (the threshold generation current
$j_{thr}\sim\frac{{1}}{{\left( {kL} \right)^{3}\left( {k\chi
_{\tau}  L} \right)^{2s}}}$, where $L$ is the length of the
interaction area).}

\section{Conclusion}

Combining the photonic crystal-based structures with vacuum
electronic devices opens the way for creation of a family of
radiation sources: volume FELs, photonic BWOs, etc.

The dispersion equations describing electron beam instability in a
bounded photonic crystal are obtained for two cases: the
conventionally considered case, when diffraction is suppressed,
and the case of direct and diffracted waves having almost equal
amplitudes. The instability law is shown to be responsible for
increase of increment of instability and decrease of length, at
which instability develops, for the case when amplitude of
diffracted wave is comparable with that of direct one that happens
in the vicinity of $\pi$-point of dispersion curve. Such
instability law enables application of photonic crystals for
development of THz sources at electron beam current densities
available at modern accelerators.

Some beneficial options are additionally available, namely: use
for radiation generation of multiple either pencil-like or sheet
electron beams instead of single annular or sheet one and, thus,
establishing the beam-wave interaction within the whole crystal
cross-section and increasing the efficiency of radiation source.

\end{document}